\documentclass[12pt]{article}  
\usepackage{hpatex}
\usepackage{graphicx}
\begin{document}
\title{Relativistically rotating dust}
\authors{Gernot Neugebauer, Andreas Kleinw\"achter and Reinhard Meinel}
\address{Max--Planck--Gesellschaft,
Arbeitsgruppe Gravitationstheorie
an der Universit\"at Jena,\\
Max--Wien--Platz 1, D--07743 Jena, Germany}
\abstract{ 
Dust configurations play an important role in astrophysics and are the
simplest models for rotating bodies. The physical properties of 
the general--relativistic global solution for the rigidly rotating disk of 
dust, which has been found recently as the solution of a boundary value 
problem, are discussed.
}
\section{Introduction}
Dust, as the simplest phenomenological material, is a good model for 
astrophysical and cosmological studies. As a source of gravitational fields, 
dust may be interpreted, in a hydrodynamical language, as a many--particle 
system, whose particles (mass--elements) interact \linebreak 
via gravitational forces 
alone. While the cosmological relevance of that model has been 
de\-mon\-stra\-ted
very early \cite{fg}, investigations of isolated dust configurations are 
rather rare \cite{os}.

This paper is meant to discuss dust as a model for rotating bodies and to
focus attention on the relativistic and ultrarelativistic behaviour of a
rotating dust cloud. For this reason, we consider the relativistic
generalization of the classical Maclaurin disk, which is the flattened
(two--dimensional) limit of the famous Maclaurin spheroids. The corresponding
global solution to the Einstein equations has recently been found as the
solution of a boundary value problem \cite{nm1}, first formulated and 
approximately solved by Bardeen and Wagoner \cite{bw1}, \cite{bw2}. In this
connection, the appearance of a boundary value problem requires a comment:
The reason for the complete absence of global solutions describing any
uniformly rotating perfect fluid ball is that there is no systematic
procedure for constructing solutions of the non--linear Einstein
equations inside the source and matching them to exterior solutions along
an unknown (`free') surface such that those global solutions are regular
everywhere. However, in the disk limit, in which the perfect fluid becomes
dust\footnote{Strictly speaking, the pressure -- to -- mass--density ratio 
tends to zero everywhere. (The maximum pressure -- at the center -- remains 
finite while the volume mass--density becomes infinite.) The radial 
distribution of the surface mass--density as well as the exterior 
gravitational field of this object become identical with those of a 
`genuine' dust disk.}, the body has no longer an interior region; it just 
consists of surface, so to say. Now, the surface conditions (vanishing 
pressure along the surface etc.) can be considered as the boundary 
conditions on a regular stationary and axisymmetric gravitational vacuum 
field (only the radius remains `free') and this fact makes disk problems 
accessible to a systematic treatment. Namely, it has been shown that 
`inverse' methods leading to linear structures  (Riemann--Hilbert problems, 
linear integral equations) apply in this case. Following this idea, it was 
possible to describe the gravitational field of the rigidly rotating disk of 
dust mentioned above in terms of two integral equations \cite{nm1}. One of 
them, the `small' one, describes the behaviour of the gravitational field 
along the axis of symmetry and the physics on the disk (fields, 
mass--density, \dots). It turned out that its solution can be represented 
in terms of elliptic functions \cite{nm2}. The other one (the `big' integral 
equation) makes use of the solution of the `small' integral equation and 
describes the gravitational field everywhere. Surprisingly, this equation 
and its corresponding
Riemann--Hilbert problem could be solved in terms of hyperelliptic
functions \cite{nm3}.

In the next section we will briefly repeat the mathematical formulation of
the boundary value problem and its solution. The main intent of this paper,
however, is the discussion of the physical properties of the disk of 
dust solution in section 3. Finally, general conclusions 
form the last section.

\section{The boundary value problem and its solution}

To describe a rigidly rotating disk of dust we use cylindrical 
Weyl--Lewis--Papapetrou coordinates
\begin{equation}
ds^2=e^{-2U}[e^{2k}(d\rho^2+d\zeta^2)+\rho^2d\varphi^2]-
e^{2U}(dt+ad\varphi)^2,
\label{line}
\end{equation}
which are adapted to stationary axisymmetric problems ($U=U(\rho,\zeta)$,
$a=a(\rho,\zeta)$, $k=k(\rho,\zeta)$). In these coordinates,
the vacuum Einstein equations are equivalent to the Ernst equation
\begin{equation}
(\Re f)(f,_{\rho \rho}+f,_{\zeta\zeta} + \frac{1}{\rho}f,_{\rho})
=f,_{\rho}^2+f,_{\zeta}^2
\label{ernst}
\end{equation}
for the complex function
\begin{equation}
f(\rho,\zeta) = e^{2U}+ib
\end{equation}
with
\begin{equation}
a,_{\rho}=\rho e^{-4U}b,_{\zeta},\quad a,_{\zeta}=-\rho e^{-4U}b,_{\rho}
\label{ab}
\end{equation}
and
\begin{equation}
k,_{\rho}=\rho[U,_{\rho}^2-U,_{\zeta}^2+
\frac{1}{4}e^{-4U}(b,_{\rho}^2-b,_{\zeta}^2)],\quad
k,_{\zeta}=2\rho(U,_{\rho}U,_{\zeta}+\frac{1}{4}e^{-4U}b,_{\rho}b,_{\zeta}).
\end{equation}
As a consequence of the Ernst equation (\ref{ernst}), the integrability
conditions $a,_{\rho \zeta}=a,_{\zeta \rho}$, $k,_{\rho \zeta}=
k,_{\zeta \rho}$
are automatically satisfied and the metric functions $a$ and $k$ may be
calculated from the Ernst potential $f$. Thus, it is sufficient to consider
the Ernst equation alone.

The metric (\ref{line}) allows an Abelian group of motions $G_2$ with the
generators (Killing vectors)
\begin{displaymath}
\xi^i=\delta^i_t,\quad \xi^i\xi_i<0 \quad \mbox{(stationarity)},
\end{displaymath}
\begin{equation}
\eta^i=\delta^i_{\varphi},\quad \eta^i\eta_i>0 \quad \mbox{(axisymmetry)},
\label{Kill}
\end{equation}
where the Kronecker symbols $\delta^i_t$ and $\delta^i_{\varphi}$ indicate
that $\xi^i$ has only a $t$--component ($\xi^t=1$) whereas $\eta^i$
points into the azimuthal $\varphi$--direction (its trajectories have
to form closed circles!). By applying (\ref{Kill}) we get from
(\ref{line}) the invariant representations
\begin{equation}
e^{2U}=-\xi_i\xi^i,\quad a=-e^{-2U}\eta_i\xi^i
\label{Ua}
\end{equation}
for the ``Newtonian'' gravitational potential $U$ and the ``gravitomagnetic''
potential $a$.

To formulate the boundary value problem for an infinitesimally thin
rigidly rotating disk of dust with a coordinate radius $\rho_0$ let us start
from a spheroid--like rotating perfect fluid configuration (Fig.~1, left)
and interpret the rigidly rotating disk of dust as an extremely
flattened limiting case of that configuration (Fig.~1, right).

\begin{center}
\includegraphics[width=0.9\textwidth]{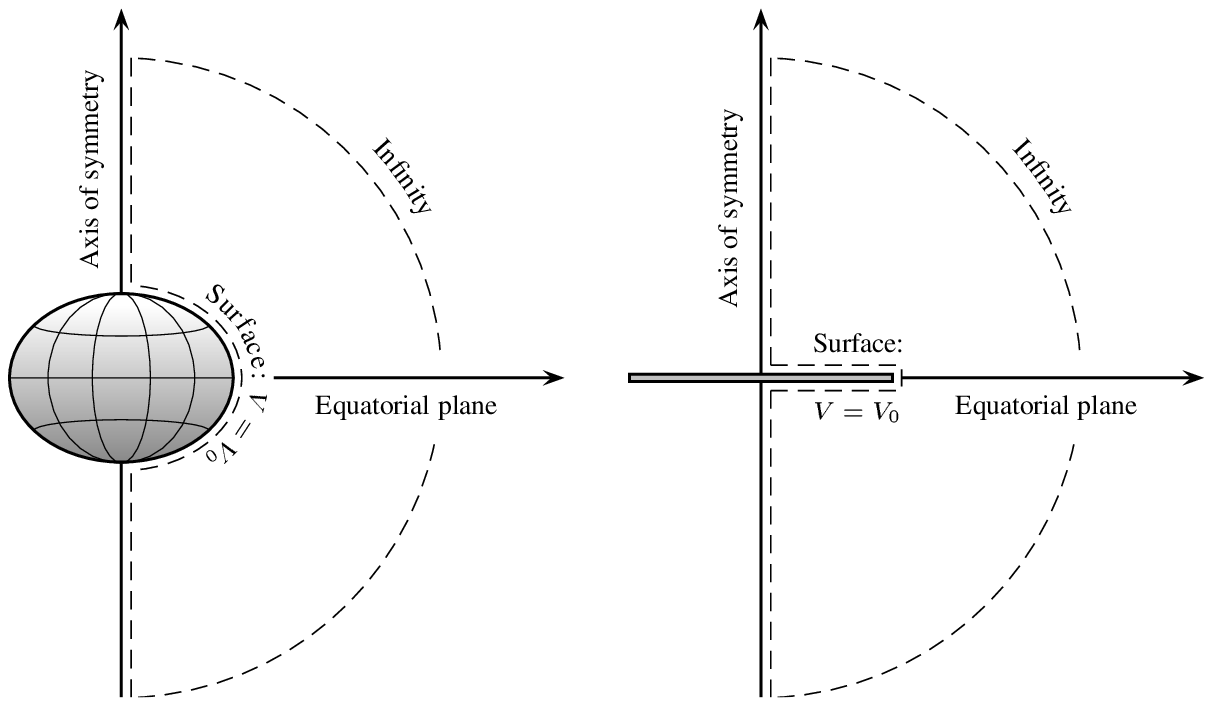}
\end{center}

\noindent {\small{\sc Fig.~1} --- The rigidly rotating disk of dust
(right--hand side) as the extremely flattened limit of a rotating perfect
fluid body (left--hand side).}

\noindent The hydrodynamics of a perfect fluid 
with  the energy--momentum tensor
\begin{equation}
T_{ik}=(\epsilon+p)u_iu_k+pg_{ik},
\label{Tik}
\end{equation}
$\epsilon$, $p$, $u_i$, $g_{ik}$ being energy density, pressure, 
four--velocity, and metric, respectively, follows from the balance equation
\begin{equation}
T^{ik};_k=0,
\label{eul}
\end{equation}
where the semicolon denotes the covariant derivative. {\it Rotational} motion
of the fluid means
\begin{equation}
u^i=e^{-V}(\xi^i+\Omega \eta^i),\quad u^iu_i=-1,
\end{equation}
i.e., the four--velocity is a linear combination of the time--like Killing 
vector $\xi^i$ and the azimuthal Killing vector $\eta^i$. Obviously,
\begin{equation}
e^{2V}=-(\xi^i+\Omega\eta^i)(\xi_i+\Omega\eta_i). 
\label{eh2V}
\end{equation}
For {\it rigidly} rotating bodies, the angular velocity
$\Omega$ is a constant,
\begin{equation}
\Omega = constant.
\end{equation}
Then, for an equation of state
\begin{equation}
\epsilon=\epsilon(p),
\label{eos}
\end{equation}
the pressure $p$ must be a function of $V$ alone,
\begin{equation}
p=p(V).
\end{equation}
This follows from Eqs.~(\ref{Tik}), (\ref{eul}).
If Eq.~(\ref{eos}) is surface--forming, the matching condition at the
surface (fluid--vacuum interface) requires
\begin{equation}
p(V_0)=0,
\end{equation}
i.e.~we have
\begin{equation}
V=V_0
\label{V0c}
\end{equation}
along the surface of the body. Eq.~(\ref{eh2V}) inspires us to introduce a
corotating frame of reference by the transformations
\begin{equation}
t'=t, \quad \varphi'=\varphi-\Omega t;
\end{equation}
\begin{equation}
\xi^{i'}=\xi^i+\Omega \eta^i, \quad \eta^{i'}=\eta^i.
\label{tr}
\end{equation}
As a consequence of Eq.~(\ref{tr}), the corotating potential $f'$ constructed
from the primed Killing vectors with the aid of the Eqs.~(\ref{Ua}) and 
(\ref{ab})
is again a solution of the Ernst equation,
\begin{equation}
(\Re f')(f',_{\rho \rho}+f',_{\zeta\zeta} + \frac{1}{\rho}f',_{\rho})
=f',_{\rho}^2+f',_{\zeta}^2.
\end{equation}
Hence, Eq.~(\ref{V0c}) tells us that the real part of the corotating Ernst 
potential is a constant along the surface of the body, $\Re f'=\exp(2V_0)$.
We may assume the validity of this result for the disk limit, too,
\begin{equation}
\left. \Re f'\right|_{\zeta=0^{\pm}}=e^{2V_0} \quad (0\le\rho\le\rho_0),
\label{f'}
\end{equation}
where $\zeta=0^{\pm}$ means top and bottom,
respectively. Thus 
we have found our first boundary condition along the disk (above and below).
As a consequence of the Einstein equations with the perfect fluid source 
(\ref{Tik}) we may conclude in the disk limit that 
\begin{equation}
(\rho^{-1}e^{4U'}a',_{\rho}),_{\rho}+(\rho^{-1}e^{4U'}a',_{\zeta}),_{\zeta}=0
\end{equation}
\newpage
\noindent 
holds everywhere {\it including} the disk. Applying the procedure known from
the transition conditions in electrodynamics, one obtains
\begin{equation}
\left. \frac{\partial a'}{\partial\zeta}\right|_{\zeta=0^+}=
\left. \frac{\partial a'}{\partial\zeta}\right|_{\zeta=0^-}
\end{equation}
at all points of the surface.
On the other hand, the reflectional symmetry of the gravitomagnetic potential
$a'$,
\begin{equation}
a'(\rho,\zeta)=a'(\rho,-\zeta)
\end{equation}
leads to
\begin{equation}
\left. \frac{\partial a'}{\partial\zeta}\right|_{\zeta=0^+}=
- \left. \frac{\partial a'}{\partial\zeta}\right|_{\zeta=0^-}
\end{equation} 
on the disk, so that $\partial a'/\partial\zeta$ has to vanish on the
disk,
\begin{equation}
\left. \frac{\partial a'}{\partial\zeta}\right|_{\zeta=0^{\pm}}=0 \quad
(0\le\rho\le\rho_0).
\end{equation}
Combining this relation with Eq.~(\ref{ab}), one has $b'=constant$ on the
disk and, after a normalization,
\begin{equation}
\left.\Im f'\right|_{\zeta=0^{\pm}}=\left.b'\right|_{\zeta=0^{\pm}}=0 \quad
(0\le\rho\le\rho_0)
\label{If'}
\end{equation}
as the other boundary condition along the disk (above and below). Since at
infinity ($\rho^2+\zeta^2\rightarrow\infty$) the space--time of any
isolated source is Minkowskian, we have to take care of 
the relation
\begin{equation}
\left. f\right|_{\rho^2+\zeta^2\rightarrow\infty}=1.
\label{A}
\end{equation}

\unitlength1cm
\begin{picture}(15,7)
\linethickness{1mm}
\put(0.5,2){\line(1,0){6}}
\thinlines
\put(6.5,2){\vector(1,0){6}}
\put(3.5,1){\vector(0,1){4}}
\put(3.4,5.4){$\zeta$}
\put(12.8,1.9){$\rho$}
\put(6.3,1.6){$\rho_o$}
\put(10.2,5){\underline{infinity:}}
\put(10.2,4.4){$f\rightarrow 1$}
\put(4,3.0){\underline{disk:}}
\put(4,2.4){$f'=e^{2V_o}$}
\end{picture}

\noindent {\small{\sc Fig.~2} --- Boundary value problem \cite{nm1}.  
$f'$ is the Ernst potential in the
corotating frame of reference defined by $\rho'=\rho$, $\zeta'=\zeta$,
$\varphi'=\varphi - \Omega t$, $t'=t$ (${u^{i}}'=e^{-V_o}{\delta^{i}_4}'$).
The solution $f(\rho,\zeta)$ has to be regular everywhere outside the disk.}

\noindent The conditions (\ref{f'}), (\ref{If'}) and (\ref{A}) 
form our system of boundary 
conditions for an extremely flattened (infinitesimally thin) rigidly
rotating disk consisting of a perfect fluid. It can be shown that the 
2--dimensional mass elements of this disk move on the geodesics of their own
field. This justifies the designation ``disk of dust'', cf.~the footnote 1.
Fig.~2 illustrates the boundary conditions on $f$. Moreover, the wanted
Ernst potential satisfying those conditions has to be regular everywhere
outside the disk. The axis of symmetry requires particular care, especially
the elementary flatness condition along the axis ($f,_{\rho}(\zeta,\rho=0)=0$). 

The boundary values depend on the two parameters $\exp(2V_0)$, 
cf.~(\ref{f'}), and $\Omega$, cf.~(\ref{tr}), so that a 2--parameter
solution can be expected from the very beginning. $\Omega$ and $\exp(2V_0)$
are ``source'' parameters with a clear physical meaning: $\Omega$ is the
constant angular velocity of the mass elements (as measured by an observer
at infinity), and $\exp(2V_0)$ defines the relative redshift
\begin{equation}
z_0=e^{-V_0}-1
\end{equation}
of a photon emitted from the center of the disk. Obviously, other parameter
pairs can be chosen. It will turn out that the `centrifugal' parameter
\begin{equation}
\mu = 2\Omega^2\rho_0^2\,e^{-2V_0}
\label{mu}
\end{equation}
is a good measure for the relativistic behaviour of the disk ($\mu$ varies 
between $\mu=0$: Minkowski space and $\mu=\mu_0=4.62966\dots$:
ultrarelativistic case). We shall present the Ernst potential $f$ in terms
of $\mu$ and $\Omega$ or $\mu$ and $\rho_0$. On the other hand, the
solution may also be characterized by the total mass $M$ and the 
$\zeta$--component of the angular momentum $J$, which far field quantities 
can be read off from a multipole expansion. Hence, a connection
\begin{equation}
\Omega=\Omega(M,J),\quad V_0 = V_0(M,J)
\end{equation}
comparable with the parameter relations of ``black hole thermodynamics''
must hold.

Let us now briefly outline the solution procedure. We have made use of the 
so--called inverse scattering method of soliton physics, which was first
utilized for the axisymmetric stationary vacuum Einstein equations by
Maison \cite{mai}, Belinski \& Zakharov \cite{bz}, Harrison \cite{ha},
Neugebauer \cite{neu}, Hauser \& Ernst \cite{haue}, Hoenselaers, Kinnersley
\& Xanthopoulos \cite{hkx}, and Aleksejew \cite{alex}. Some of these authors 
followed the line of Geroch \cite{ger}, Kinnersley \cite{ki}, Kinnersley
\& Chitre \cite{kc}, and Herlt \cite{he}. We have applied a local version
\cite{neux} in which the Ernst equation (\ref{ernst}) is the integrability
condition of the ``linear problem''
\begin{equation}
\Phi,_z=\left\{ \left( \begin{array}{cc} C & 0 \\ 0 & D \end{array} \right)
        +\lambda \left( \begin{array}{cc} 0 & C \\D & 0 \end{array} 
        \right) \right\}\Phi,
\label{Lin1}
\end{equation}
\begin{equation}
\Phi,_{\bar{z}}=\left\{ \left( \begin{array}{cc} \bar{D} & 0 \\ 0 & \bar{C} 
         \end{array} \right)
        +\frac{1}{\lambda} \left( \begin{array}{cc} 0 & \bar{D} \\ \bar{C} & 0 
        \end{array} \right) \right\}\Phi,
\label{Lin2}
\end{equation}
where $\Phi(z,\bar{z},\lambda)$ is a $2\times2$ matrix depending on the
spectral parameter
\begin{equation}
\lambda=\sqrt{\frac{K-i\bar{z}}{K+iz}} \quad  (K\,\, \mbox{a complex constant)}
\label{lam}
\end{equation}
as well as on the coordinates $z=\rho+i\zeta$, $\bar{z}=\rho-i\zeta$,
whereas $C$, $D$ and the complex conjugate quantities $\bar{C}$, $\bar{D}$ are
functions of $z$, $\bar{z}$ ($\rho$, $\zeta$) alone. Indeed, from
$\Phi,_{z\bar{z}}=\Phi,_{\bar{z}z}$ and the formulae
\begin{equation}
\lambda,_z=\frac{\lambda}{4\rho}(\lambda^2-1), \quad \lambda,_{\bar{z}}=
\frac{1}{4\rho\lambda}(\lambda^2-1)
\end{equation}
it follows that a certain matrix polynomial in $\lambda$ has to vanish.
This yields the set of first order differential equations
\begin{equation}
C,_{\bar{z}}=C(\bar{D}-\bar{C})-\frac{1}{4\rho}(C+\bar{D}),\quad
D,_{\bar{z}}=D(\bar{C}-\bar{D})-\frac{1}{4\rho}(D+\bar{C})
\label{syst}
\end{equation}
plus the complex conjugate equations. The system (\ref{syst}) has the ``first
integrals'' 
\begin{equation}
D=\frac{f,_z}{f+\bar{f}}, \quad C=\frac{\bar{f}_z}{f+\bar{f}}.
\end{equation}
Eliminating $C$ and $D$ in (\ref{syst}) one arrives at the Ernst equation 
(\ref{ernst}). Vice versa, if $f$ is a solution to the Ernst equation, 
the matrix $\Phi$ calculated from (\ref{Lin1}), 
(\ref{Lin2}) does not depend on the path of integration. The 
idea of the ``inverse methods'' is to discuss $\Phi$, 
for fixed values of $z$, $\bar{z}$,
as a holomorphic function of $\lambda$ and 
to calculate $C$ and $D$ from $\Phi$
afterwards. This is an `inverse' procedure compared 
with the `normal' way which
consists of the solution of differential equations with given coefficients. 
The term `scattering' comes from the solution 
technique of the Korteweg--de Vries
equation developed by Gardner, Greene, Kruskal \& Miura \cite{ggkm} whose
linear problem has partially the form of the (time--independent) Schr\"odinger
equation. In this case, the calculation of the coefficients consists of the
construction of the Schr\"odinger potential from the scattering data.   

To construct $\Phi$ as a function of $\lambda$ we 
have integrated the linear problem
(\ref{Lin1}), (\ref{Lin2}) along the dashed line in Fig.~1 
(right--hand side) and exploited the 
information of $C$, $D$ and $\lambda$ along 
the axis of symmetry ($f,_{\rho}=0$,
$\lambda=\pm1$), the boundary conditions on the disk which simplify $C'$ and
$D'$ (i.e.~we had here to switch to the corotating system) and the simple
structure of the linear problem at infinity ($C=D=0$).
In this way, we could pick up enough information to construct 
$\Phi(z,\bar{z},\lambda)$ completely. 
(The crucial steps were the formulation and solution of a matrix 
Riemann--Hilbert
problem in the complex $\lambda$--plane.) The linear problem (\ref{Lin1}), 
(\ref{Lin2}) tells
us that $\Phi$ at $\lambda=1$ may be normalized in a very simple way,
\begin{equation}
\Phi(z,\bar{z},\lambda=1) = 
 \left( \begin{array}{cr} \bar{f} & 1 \\ f & -1 
         \end{array} \right).
\end{equation}         
Hence, once $\Phi$ is known, the Ernst potential $f$ can be read off from
$\Phi$ ($f=\Phi_{21}(z,\bar{z},1)$).

The result for our problem is \cite{nm3}
\begin{equation}
f = \exp
\left\{\mu\left[\,\int\limits_{X_1}^{X_a}\frac{X^2\,dX}{W} +  
\int\limits_{X_2}^{X_b}\frac{X^2\,dX}{W} - 
\int\limits_{-i}^{i}\frac{hX^2\,dX}{W_1}\right]\right\},
\label{fint}
\end{equation}
where the lower integration limits $X_1$, $X_2$ are given by
\begin{equation}
X_1^2 = \frac{i - \mu}{\mu}, \quad  X_2^2 = -\frac{i + \mu}{\mu}
\quad (\Re X_1 < 0, \quad \Re X_2 > 0).
\end{equation}
whereas the upper limits $X_a$, $X_b$ must be calculated from the
integral equations
\begin{equation}
\int\limits_{X_1}^{X_a}\frac{dX}{W} + \int\limits_{X_2}^{X_b}\frac{dX}{W} = 
\int\limits_{-i}^i \frac{h dX}{W_1},
\qquad \int\limits_{X_1}^{X_a}\frac{X\,dX}{W} + 
\int\limits_{X_2}^{X_b}\frac{X\,dX}{W} = 
\int\limits_{-i}^i \frac{h X dX}{W_1}.
\label{jacobi}
\end{equation}

\noindent
Here we have introduced the abbreviations
\begin{equation}
W = W_1 W_2, \qquad W_1 = \sqrt{(X-\zeta/\rho_0)^2 +
                              (\rho/\rho_0)^2}, \qquad
W_2 = \sqrt{1 + \mu^2(1+X^2)^2}
\end{equation}
and
\begin{equation}
h = \frac{\ln\left(\sqrt{1 + \mu^2(1+X^2)^2} + \mu(1+X^2)\right)}
         {\pi i \sqrt{1 + \mu^2(1+X^2)^2} } \,.
\end{equation}
The third integral in (\ref{fint}) as well as the integrals on the 
right-hand sides in (\ref{jacobi}) have to be taken 
along the imaginary axis in the complex X-plane
with $h$ and and $W_1$ fixed according to $\Re W_1<0$ (for $\rho,\zeta$
outside the disk) and $\Re h = 0$\,.
The task of calculating the upper limits $X_a$, $X_b$ in (2.39) from
\begin{equation}
u = \int\limits_{-i}^i \frac{h\, dX}{W_1} \,, \quad
v = \int\limits_{-i}^i \frac{h X dX}{W_1}
\label{uv}
\end{equation}
is known as Jacobi's famous inversion problem. 
G\"opel \cite{goe} and Rosenhain
\cite{ro} were able to express the hyperelliptic functions $X_a(u,v)$ and
$X_b(u,v)$ in terms of (hyperelliptic) theta functions. Later on it turned
out that even the first two integrals in (\ref{fint}) can be expressed by
theta functions in $u$ and $v$! A detailed introduction
into the related mathematical theory which was founded by Riemann  
and Weierstra\ss\  may be found in \cite{st}, \cite{kra}, 
\cite{bob}.
The representation of the Ernst potential (\ref{fint}) 
in terms of theta functions can
be found in Stahl's book, see \cite{st}, page 311, Eq.~(5). Here is the
result: Defining a theta function
$\vartheta(x,y;p,q,\alpha)$
by\footnote{We use
$\vartheta = \vartheta \left[ \begin{array}{cc}   0 & 0 \\ 1 & 1
                        \end{array} \right] $
where the bracket indicates the characteristic, see \cite{kra}.}
\begin{equation}
\vartheta(x,y;p,q,\alpha) = \sum\limits_{m=-\infty}^{\infty}
                         \sum\limits_{n=-\infty}^{\infty}
(-1)^{m+n} p^{m^2} q^{n^2} e^{2mx + 2ny + 4m n \alpha}
\label{theta}
\end{equation}
one can reformulate the expressions (\ref{fint}), (\ref{jacobi}) to give
\begin{equation}
f = \frac{ \vartheta(\alpha_0 u + \alpha_1 v - C_1,
                  \beta_0 u + \beta_1 v - C_2;     p,q,\alpha) }
         { \vartheta(\alpha_0 u + \alpha_1 v + C_1,
                  \beta_0 u + \beta_1 v + C_2;     p,q,\alpha) } \,\,
    e^{-(\gamma_0 u + \gamma_1 v + \mu w)}
\label{ftheta}
\end{equation}
with $u$ and $v$ as in (\ref{uv}) and
\begin{equation}
w = \int\limits_{-i}^i \frac{h X^2 dX}{W_1} \,.
\end{equation}
The normalization parameters $\alpha_0$, $\alpha_1$; $\beta_0$, $\beta_1$;
$\gamma_0$, $\gamma_1$, the moduli $p$, $q$, $\alpha$
of the theta function and the quantities $C_1$, $C_2$ are defined on the two
sheets of the hyperelliptic Riemann surface related to
\begin{equation}
W = \mu \sqrt{(X-X_1)(X-\bar{X_1})(X-X_2)(X-\bar{X_2})
              (X-i\bar{z}/\rho_0)(X+i z/\rho_0)},
\end{equation} 
see Figure 3.

\begin{center}
\includegraphics[width=0.7\textwidth]{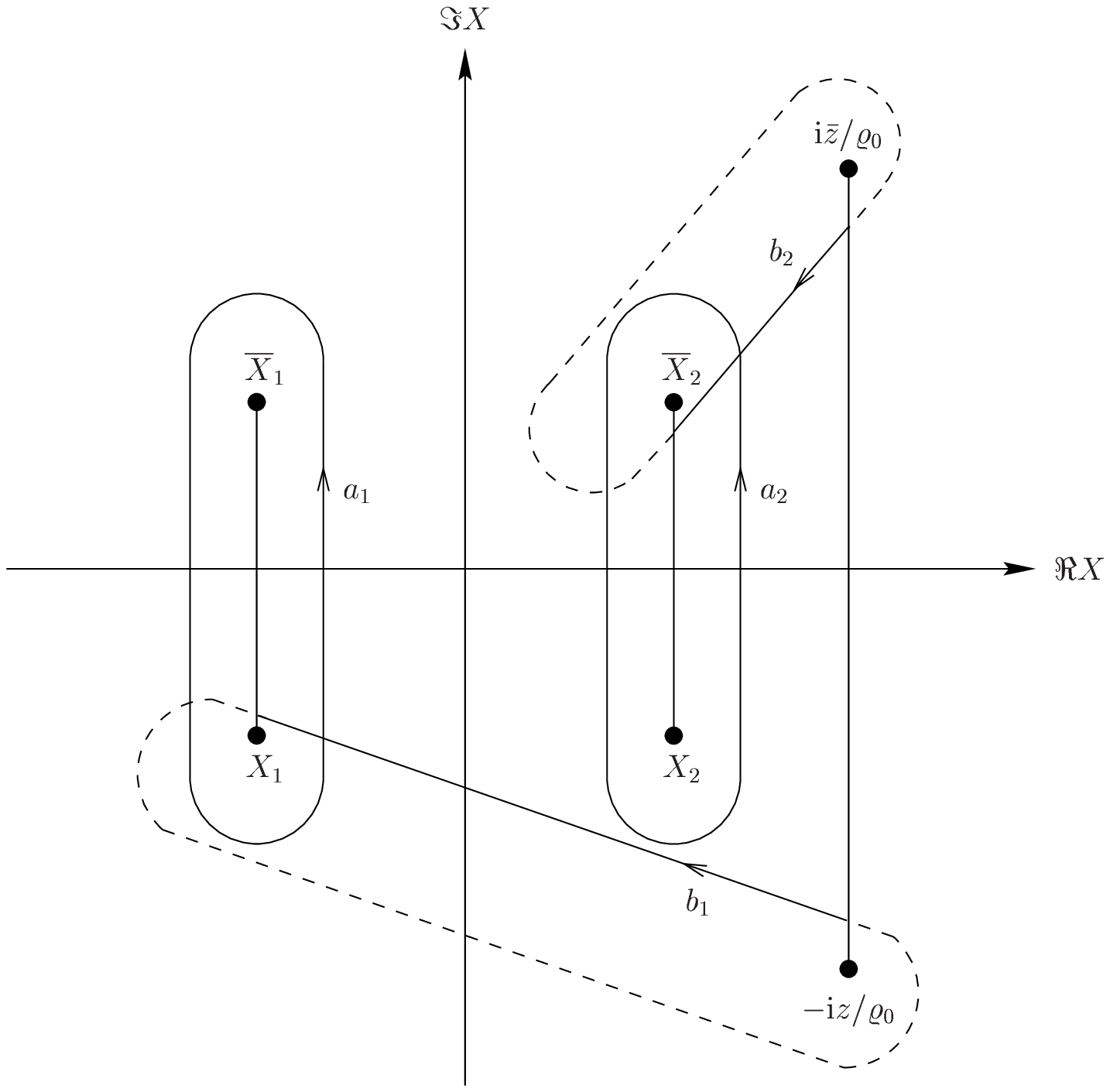}
\end{center}

\noindent {\small{\sc Fig.~3} --- Riemann surface with cuts between the 
branch points $X_1$ and
$\bar{X}_1$, $X_2$ and $\bar{X}_2$, $-iz/\rho_0$ and $i\bar{z}/\rho_0$.
Also shown are the four periods $a_i$ and $b_i$ ($i=1,2$). 
(Continuous/dashed lines 
belong to the upper/lower sheet defined by $W\rightarrow\pm\mu X^3$ as 
$X\rightarrow\infty$.)}

\noindent There are two normalized Abelian differentials
of the first kind
\begin{eqnarray}
d\omega_1 & = & \alpha_0 \frac{d X}{W} + \alpha_1 \frac{X dX}{W} \\
d\omega_2 & = & \beta_0 \frac{d X}{W} + \beta_1 \frac{X dX}{W}
\end{eqnarray}
defined by
\begin{equation}
\oint\limits_{a_m} d\omega_n = \pi i\, \delta_{mn} \, \quad (m=1,2;\, 
n=1,2)\,.
\label{aper}
\end{equation}
Eq.~(\ref{aper}) consists of four linear algebraic equations 
and yields the four
parameters $\alpha_0$, $\alpha_1$, $\beta_0$, $\beta_1$ in terms of integrals
extending over the closed (deformable) curves $a_1$, $a_2$. It can be shown
that there is one normalized Abelian differential of the third kind
\begin{equation}
d\omega = \gamma_0 \frac{d X}{W} + \gamma_1 \frac{X dX}{W}
              + \mu \frac{X^2 dX}{W}
\end{equation}
with vanishing $a-$periods,
\begin{equation}
\oint\limits_{a_j} d\omega = 0 \,\quad (j=1,2)\,.
\end{equation}
This equation defines $\gamma_0$, $\gamma_1$ (again via a linear algebraic
system). The Riemann matrix
\begin{equation}
(B_{i j}) = \left( \begin{array}{cc}
                     \ln p & 2\alpha \\
                     2\alpha & \ln q
                   \end{array} \right) \, \quad (i=1,2;\, j=1,2)
\end{equation}
(with negative definite real part) is given by
\begin{equation}
B_{i j} = \oint\limits_{b_i} d\omega_j
\end{equation}
and defines the moduli $p$, $q$, $\alpha$ of the theta function (\ref{theta}).
Finally, the quantities $C_1$, $C_2$ can be calculated by
\begin{equation}
C_i = -\int\limits_{-iz/\rho_0}^{\infty^+} d\omega_i \, \quad (i=1,2) \,,
\end{equation}
where $+$ denotes the upper sheet.
Obviously, all the quantities entering the theta functions and the
exponential function in (\ref{ftheta}) 
can be expressed in terms of well--defined
integrals and depend on the three parameters $\rho/\rho_0$,
$\zeta/\rho_0$, $\mu$.
The corresponding ``tables'' for $\alpha_i$, $\beta_i$, $\gamma_i$, 
$C_i$, $B_{i j}$,
$u$, $v$, $w$ can easily be calculated by numerical integrations. Fortunately,
theta series like (\ref{theta}) converge rapidly. For $0<\mu<\mu_0$,
the solution (\ref{ftheta}) is analytic {\it everywhere} outside the disk --
even at the rings $-iz/\rho_0=X_1$, $X_2$.

Figures 4 and 5 give an impression of the Ernst potential for $\mu=3$, as an
example.

\begin{center}
\includegraphics[width=0.6\textwidth]{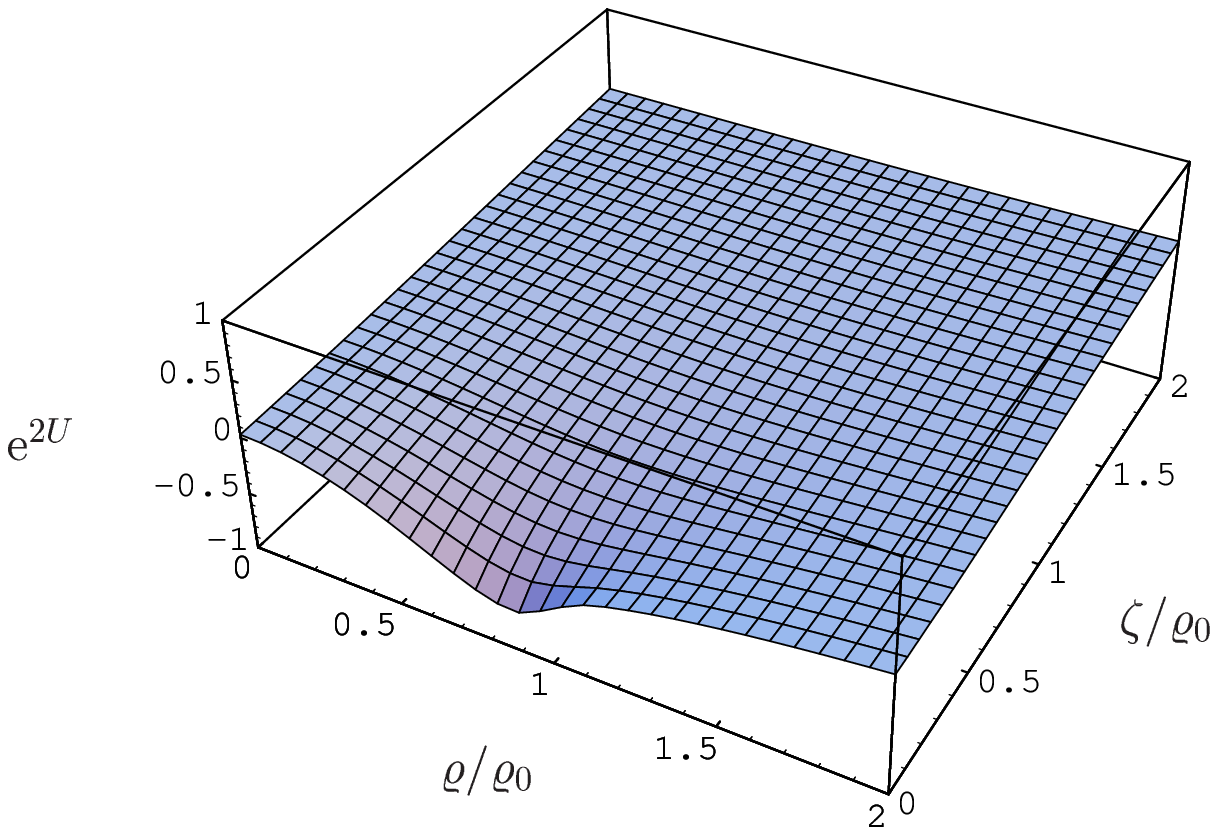}
\end{center}
\noindent {\small{\sc Fig.~4} --- The real part ($e^{2U}$) of the 
Ernst potential for $\mu = 3$.}

\begin{center}
\includegraphics[width=0.6\textwidth]{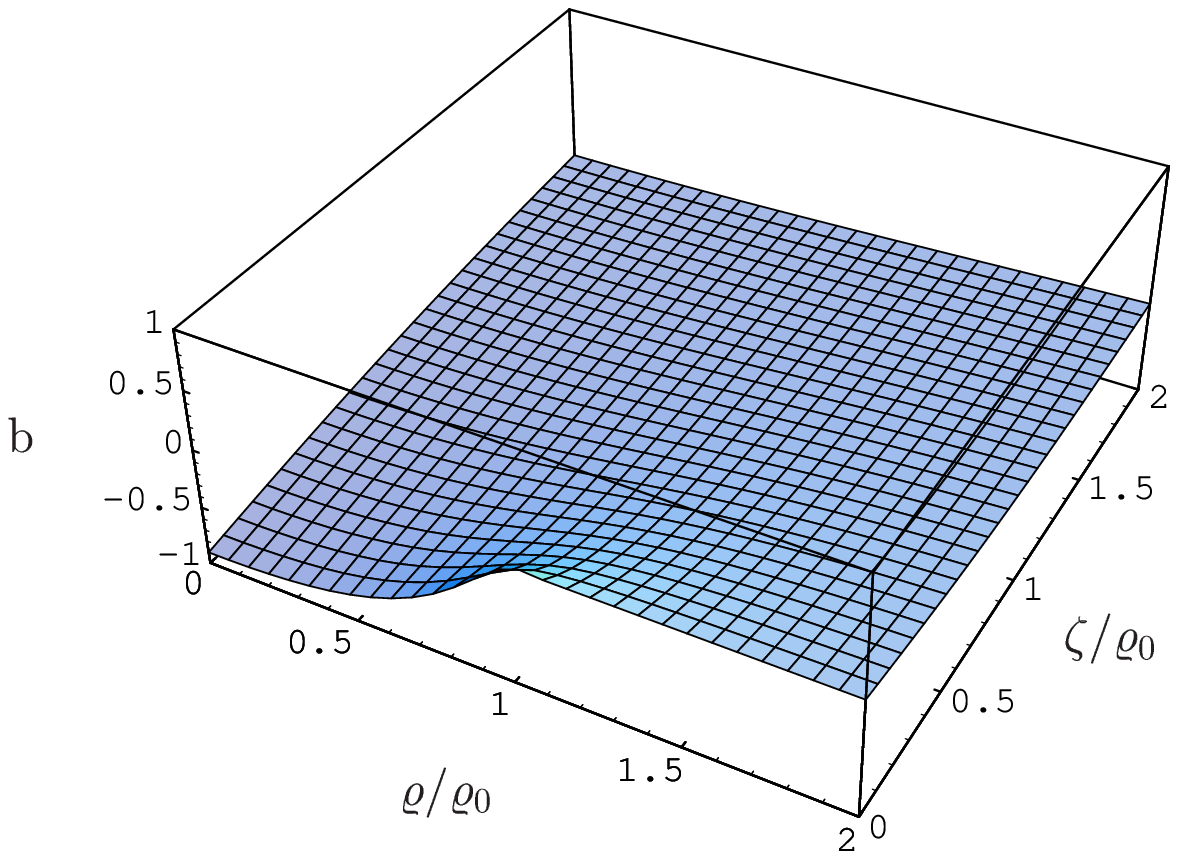}
\end{center}
\noindent {\small{\sc Fig.~5} --- The imaginary part ($b$) of the 
Ernst potential for $\mu = 3$.}

\section{Physical properties}
In dimensionless coordinates, $\rho/\rho_0$, $\zeta/\rho_0$, the solution
depends on the single parameter $\mu$. The original 
parameter $V_0$ entering our
boundary value problem can be calculated from $\mu$ via 
$\Re f(\rho=0,\zeta=0^+)$
(on the axis $\rho = 0$ we have $\Re f' = \Re f$) leading to \cite{nm2}
\begin{equation}
V_0 = -\, \frac{1}{2} \sinh ^{-1} \left\{\mu + 
\frac{1+\mu^2}{\wp[I(\mu);\frac{4}{3}\mu^2-4,
\frac{8}{3}\mu(1+\frac{\mu^2}{9})]
-\frac{2}{3}\mu} \right\},
\label{V0}
\end{equation} 
\begin{equation}
I(\mu) = \frac{1}{\pi}\int_0^{\mu}\,\frac{\ln(x+\sqrt{1+x^2})dx}
{\sqrt{(1+x^2)(\mu-x)}},
\end{equation}
where $\wp$ is the Weierstra\ss\ function defined by
\begin{equation}
\int_{\wp(x;g_2,g_3)}^{\infty}\,\frac{dt}{\sqrt{4t^3-g_2t-g_3}} = x.
\end{equation}
Fig.~6 shows $\exp[2V_0(\mu)]$ in the range $0<\mu<\mu_0=4.62966184\dots$ with
$\mu_0$ being the first zero of the denominator in (\ref{V0}). 
This corresponds
to $0>V_0>-\infty$, where $|V_0|\ll 1$ is the Newtonian limit. 
Note that, according to (\ref{mu}) and (\ref{V0}), $\Omega\rho_0$ is a given 
function of $\mu$, and we can use either $\rho_0$ or $\Omega$ as the second
parameter.
\newpage


\begin{center}
\includegraphics[width=0.5\textwidth]
   {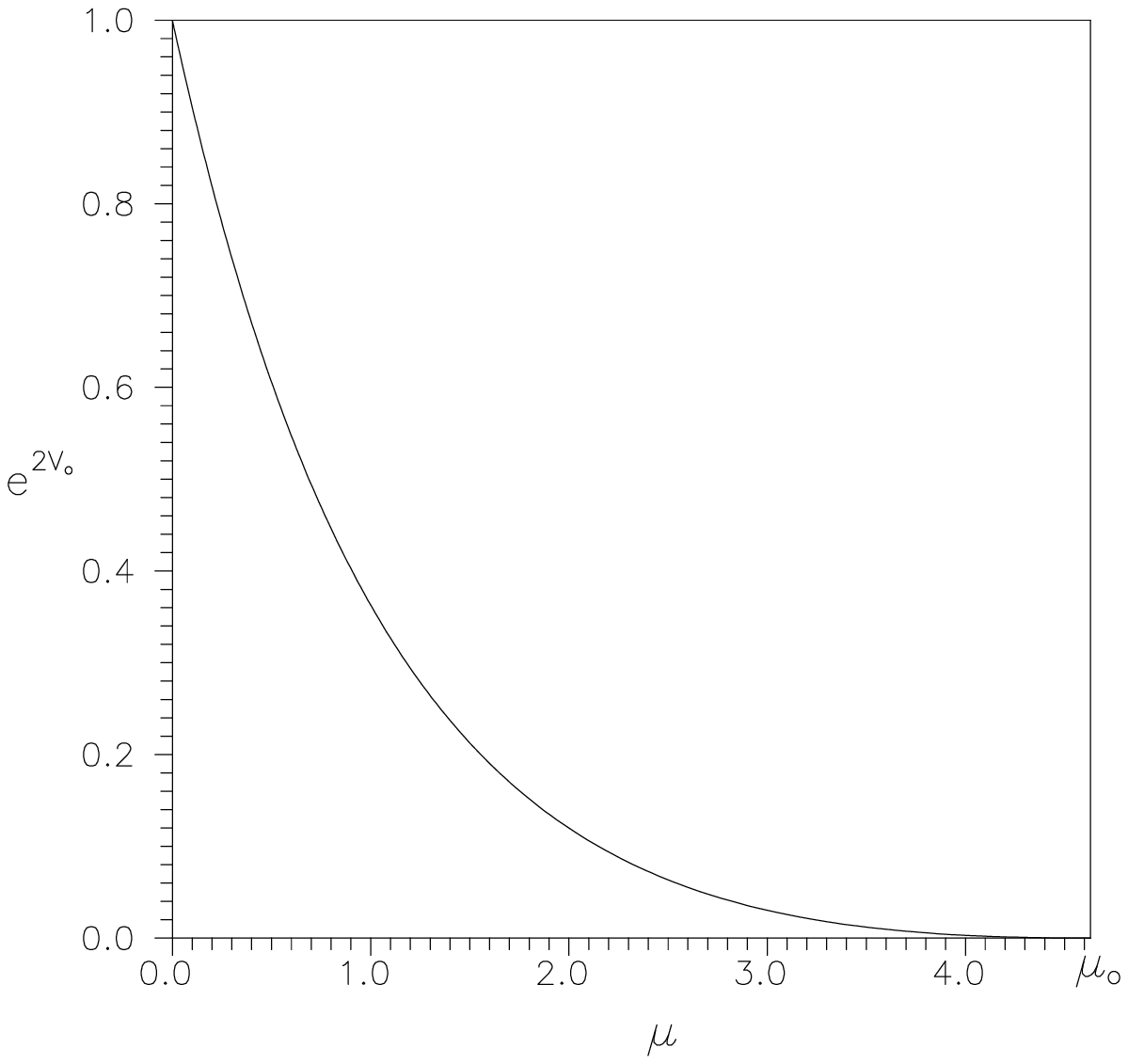}
\end{center}   

\noindent {\small{\sc Fig.~6} --- The corotating 
Ernst potential $f'=e^{2V_0}$
in the disk as a function of the parameter $\mu$.}
 
\noindent In Fig.~7 the invariant baryonic (proper) 
surface mass--density $\sigma_p$ can 
be found as a function of $\rho/\rho_0$ for several values of $\mu$. 
Note that 
the volume mass--density $\epsilon$ entering the energy--momentum tensor
($T_{ik} = \epsilon u_iu_k$) may be expressed formally by
\begin{equation}
\epsilon = \sigma_p(\rho) e^{U-k} \delta(\zeta),
\end{equation}
where $\delta(\zeta)$ is the usual Dirac delta--distribution. 


\begin{center}
\includegraphics[width=0.5\textwidth]
   {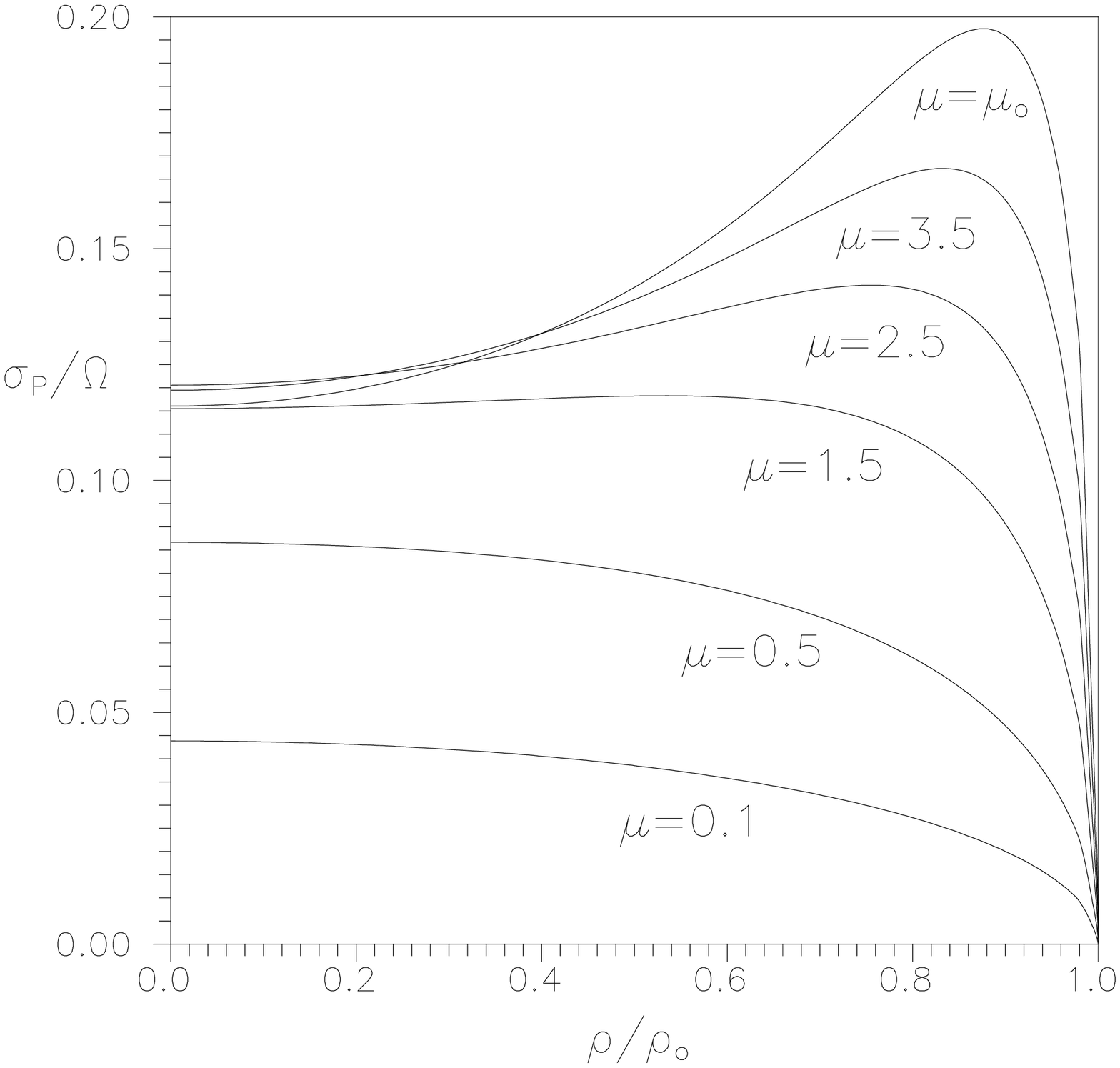}
\end{center}

\noindent {\small{\sc Fig.~7} --- The surface mass--density $\sigma_p$. The
normalized quantity $\sigma_p/\Omega$ is shown as a function of 
$\rho/\rho_0$.}

\noindent $\sigma_p$ can be calculated from
\begin{equation}
\sigma_p = \left. \frac{1}{2\pi}\,e^{U-k}\,\frac{\partial U'}{\partial \zeta}
\right|_{\zeta=0^+}.
\end{equation}   
A more explicit expression has been given in \cite{nm2}.
From $\sigma_p$ we can calculate the total baryonic mass $M_0$, the
gravitational mass $M$, and the total angular momentum $J$:
\begin{equation}
M_0 = \int\limits_{\Sigma} \epsilon\sqrt{-g}\,u^4 d^3x = 
2\pi \,e^{-V_0}\int\limits_0^{\rho_0}\sigma_p e^{k-U}\rho d\rho,
\end{equation}
\begin{equation}
M =  2\int\limits_{\Sigma} (T_{ab}-\frac{1}{2}Tg_{ab})n^a\xi^b dV =
2\pi\int\limits _0^{\rho_0}\sigma_p e^{k-U}\rho d\rho + 4\pi\Omega e^{-V_0}
\int\limits_0^{\rho_0} \sigma_p e^{k-U} u^i\eta_i \rho\,d\rho,
\end{equation}
\begin{equation}
J = - \int\limits_{\Sigma} T_{ab} n^a\eta^b dV = 2\pi \,e^{-V_0}
\int\limits _0^{\rho_0}\sigma_p e^{k-U} u^i\eta_i \rho\,d\rho,
\end{equation}
where $\Sigma$ is the spacelike hypersurface $t=constant$ with the unit 
future--pointing normal vector $n^a$. Note that $M=\exp(V_0)M_0 + 2\Omega J$,
cf.~\cite{nh}.
Alternatively, $M$ and $J$ -- as the first gravitational multipole moments --
may be obtained from the asymptotic expansion of the Ernst potential, e.g.~on
the symmetry axis. The dependence of $M_0$, $M$ and $J$ on $\mu$ can be seen
in Fig.~8. 


\begin{center}
\includegraphics[width=0.5\textwidth]
   {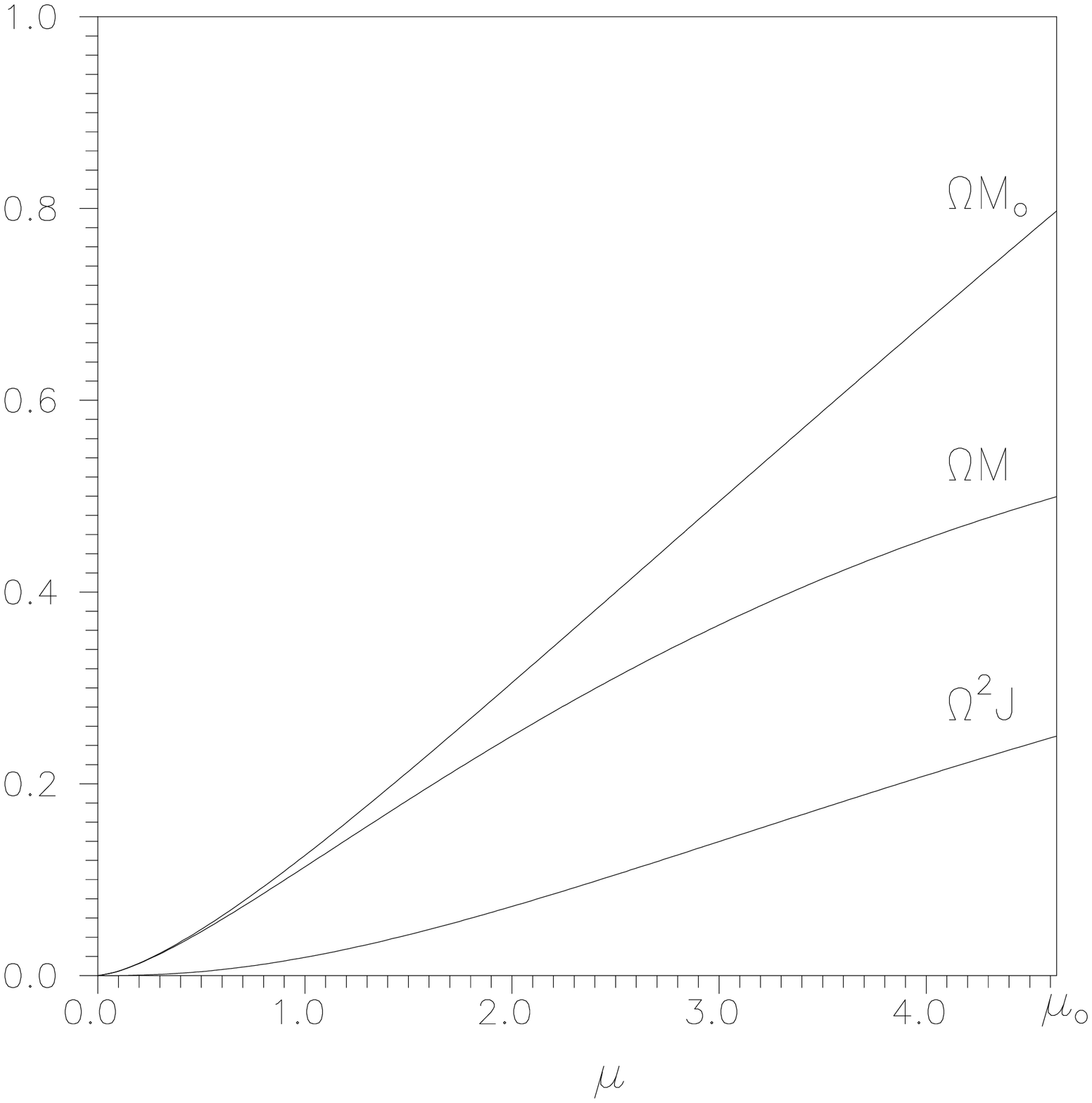}
\end{center}   

\noindent {\small{\sc Fig.~8} --- Baryonic mass $M_0$, gravitational mass
$M$ and angular momentum $J$. The normalized quantities $\Omega M_0$, 
$\Omega M$ and $\Omega^2 J$ are shown in dependence on $\mu$.}

\noindent The relative binding energy $(M_0-M)/M_0$ as well as the 
characteristic quantity $M^2/J$ are shown in Fig.~9.

\newpage

\begin{center}
\includegraphics[width=0.5\textwidth]
   {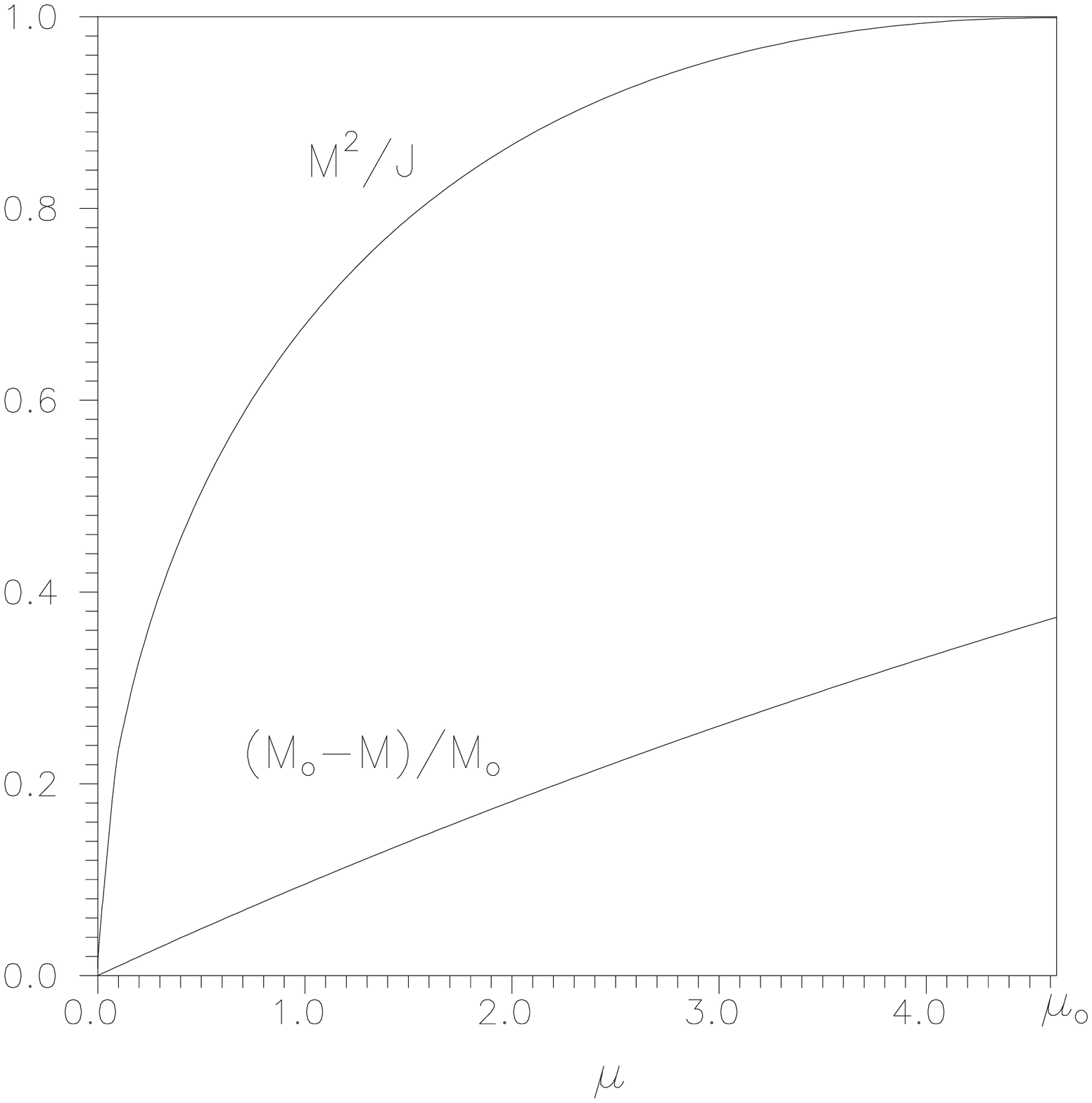}
\end{center}   

\noindent {\small{\sc Fig.~9} --- Relative binding energy and $M^2/J$.}

\noindent For $\mu\rightarrow\mu_0$, $M$ and $J$
together with all the other multipole moments approach exactly the values of
the extreme Kerr solution ($\Omega=1/\,2M$ may be identified with the angular 
velocity of the horizon) \cite{kmn}. In fact, the solution becomes identical 
with the extreme Kerr solution for all values of $\rho$ and $\zeta$ except
$\rho=\zeta=0$, which represents the horizon of the extreme Kerr black hole.
Note that, for non--vanishing $\Omega$ (finite $M$), $\rho_0\rightarrow 0$
as $\mu\rightarrow\mu_0$. 

\begin{center}
\includegraphics[width=0.5\textwidth]
   {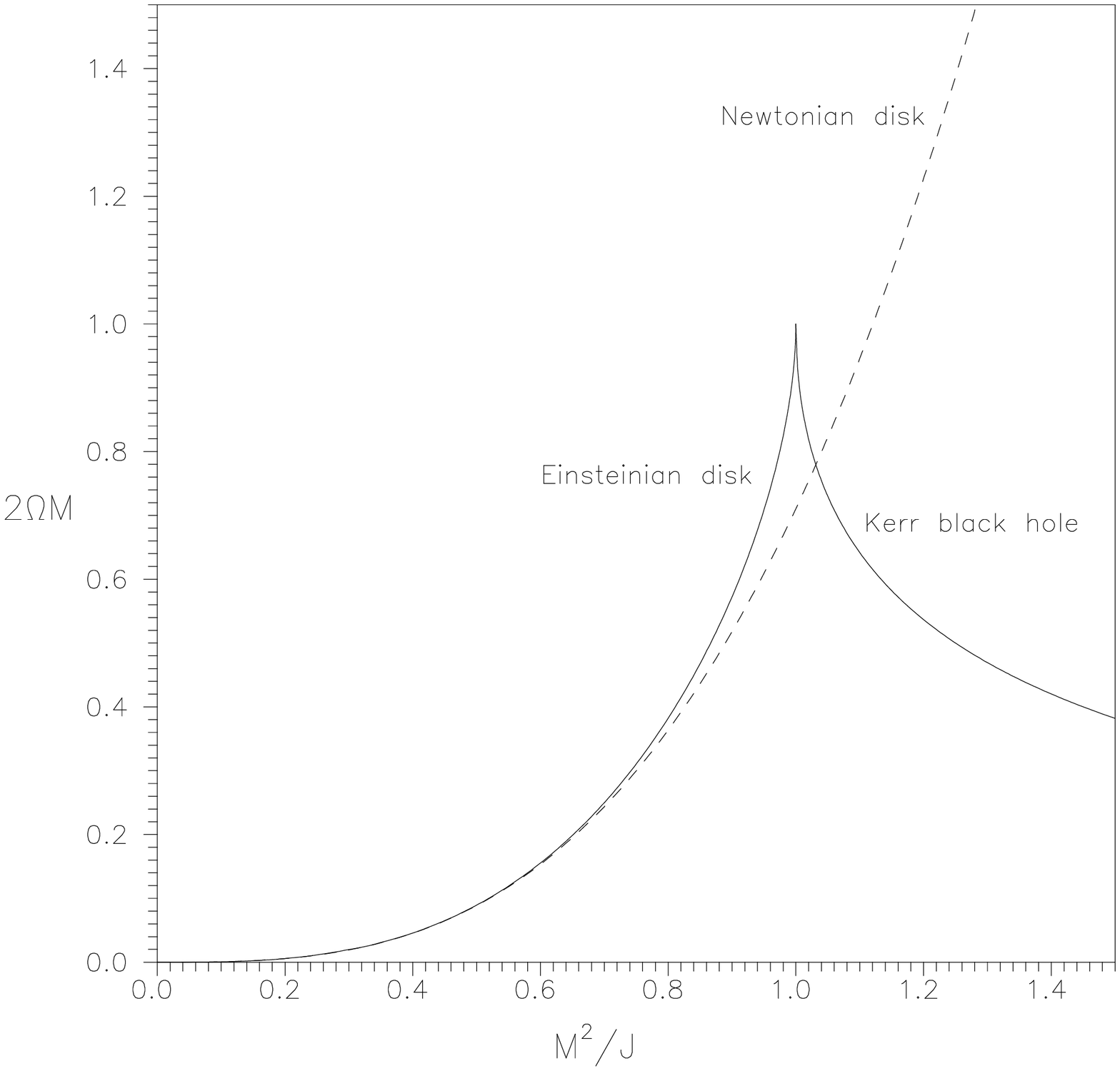}
\end{center}   

\noindent {\small{\sc Fig.~10} --- Relation between $\Omega M$ and $M^2/J$
for the classical Maclaurin disk (dashed line), the general--relativistic
dust disk and the Kerr black--hole \cite{nm1}.}

\noindent Fig.~10 combines the parameter relations between $\Omega M$ 
and $M^2/J$ for
the dust disk as well as for the Kerr black--hole. Both branches are
connected at the point $M^2/J=1$.

Another limit of the space--time for
$\mu\rightarrow\mu_0$ is obtained for finite values of $\rho/\rho_0$ and
$\zeta/\rho_0$, see also \cite{bw2}. The interpretation of the solution for 
$\mu>\mu_0$ is beyond the scope of this paper. We only want to mention here,
that there are further zeros $\mu_n$ ($n=1,2,\dots$) of the denominator in
(\ref{V0}) leading always to the extreme Kerr metric ($\mu_1=38.70908\dots$,
$\mu_2=176.92845\dots$, \dots). 

A characteristic feature of relativistically rotating bodies are
dragging effects due to the gravitomagnetic potential. Dragging effects near 
the rigidly rotating disk of dust have been discussed in \cite{mk}. In
particular, the dust disk generates an ergoregion for $\mu>\mu_e=1.68849\dots$,
see Fig.~11. In this region the Killing vector $\xi^i$ becomes spacelike. As a
consequence, $d\varphi/dt > 0$ must hold for any timelike worldline there.
Thus, seen from infinity, any observer inside the ergoregion is forced to
rotate in the same direction as the disk. For $\mu\rightarrow\mu_0$ the 
well--known ergosphere of the extreme Kerr black hole appears.


\begin{center}
\includegraphics[width=0.3\textwidth]
   {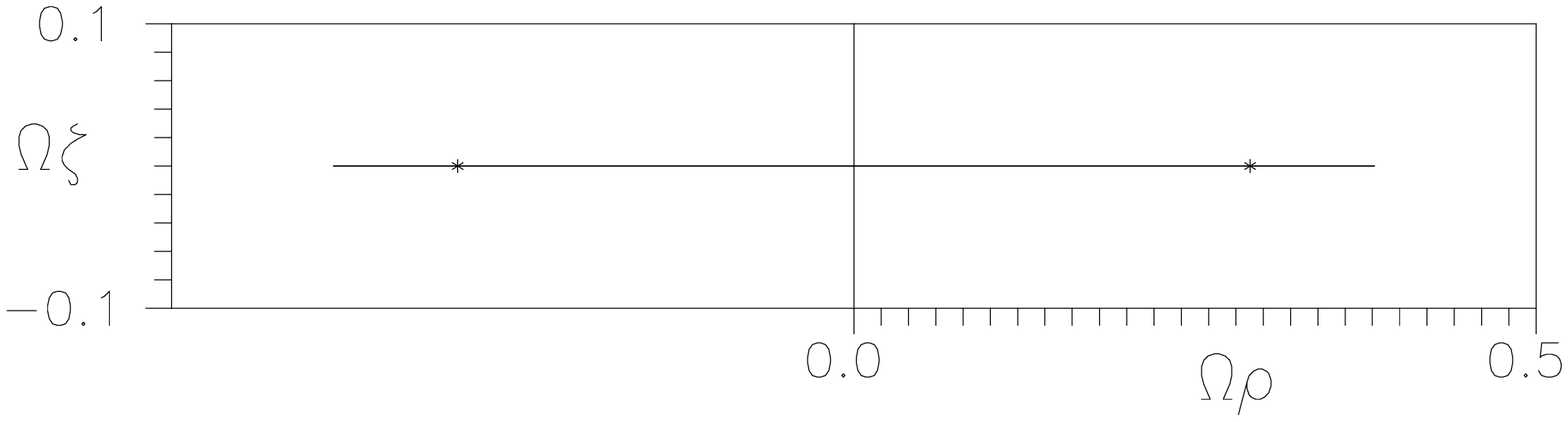}{\qquad $\mu = \mu _{\rm e}$}
\end{center}   
\begin{center}
\includegraphics[width=0.3\textwidth]
   {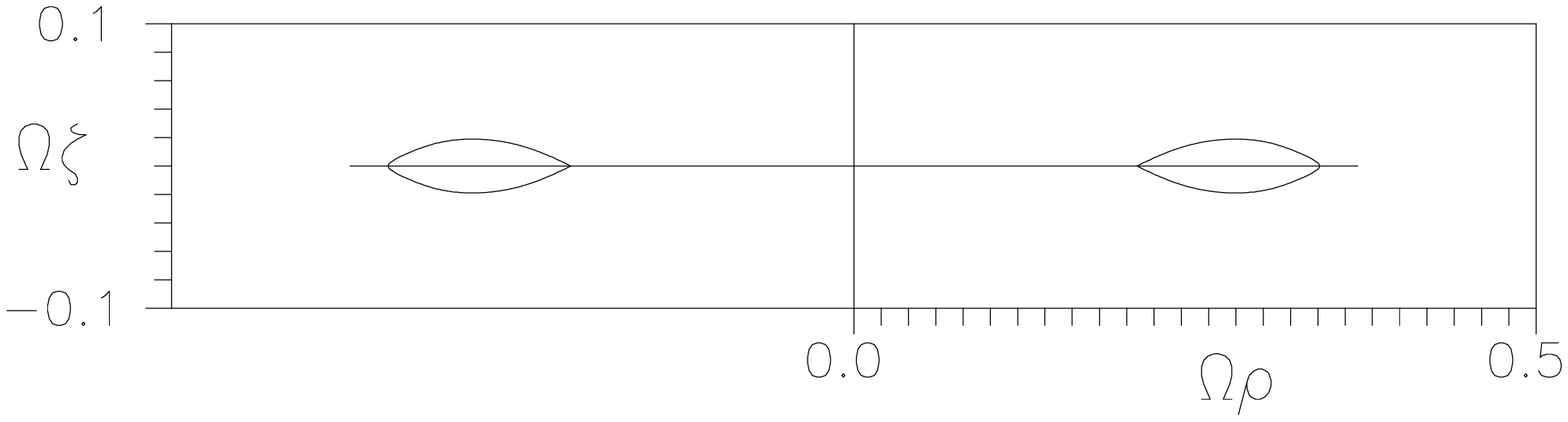}{\qquad $\mu = 1.8$}
\end{center}   
\begin{center}
\includegraphics[width=0.3\textwidth]
   {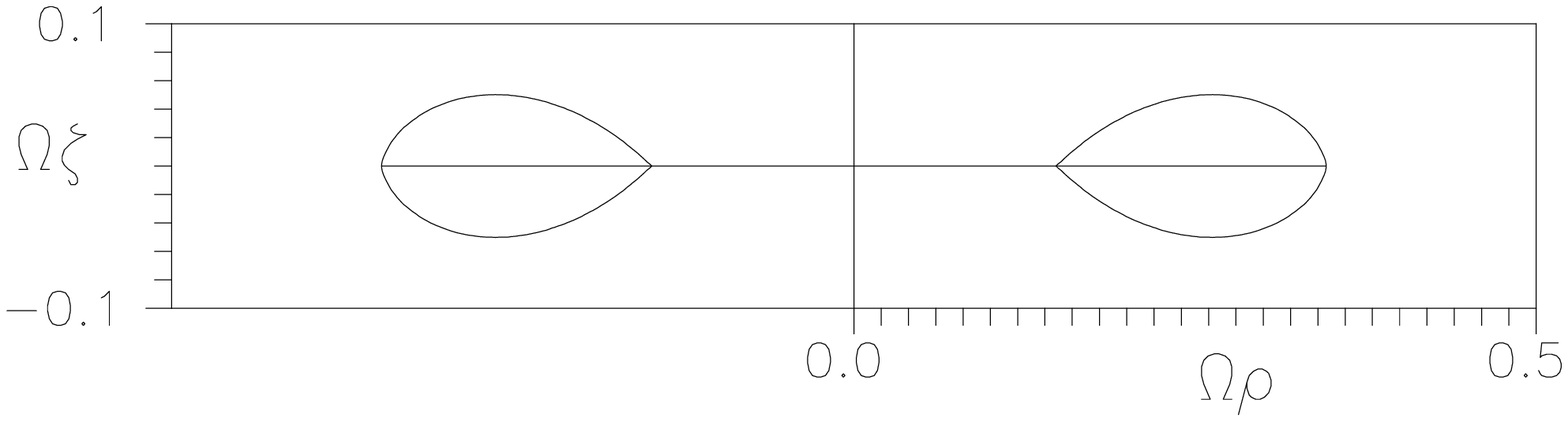}{\qquad $\mu = 2.0$}
\end{center}   
\begin{center}
\includegraphics[width=0.3\textwidth]
   {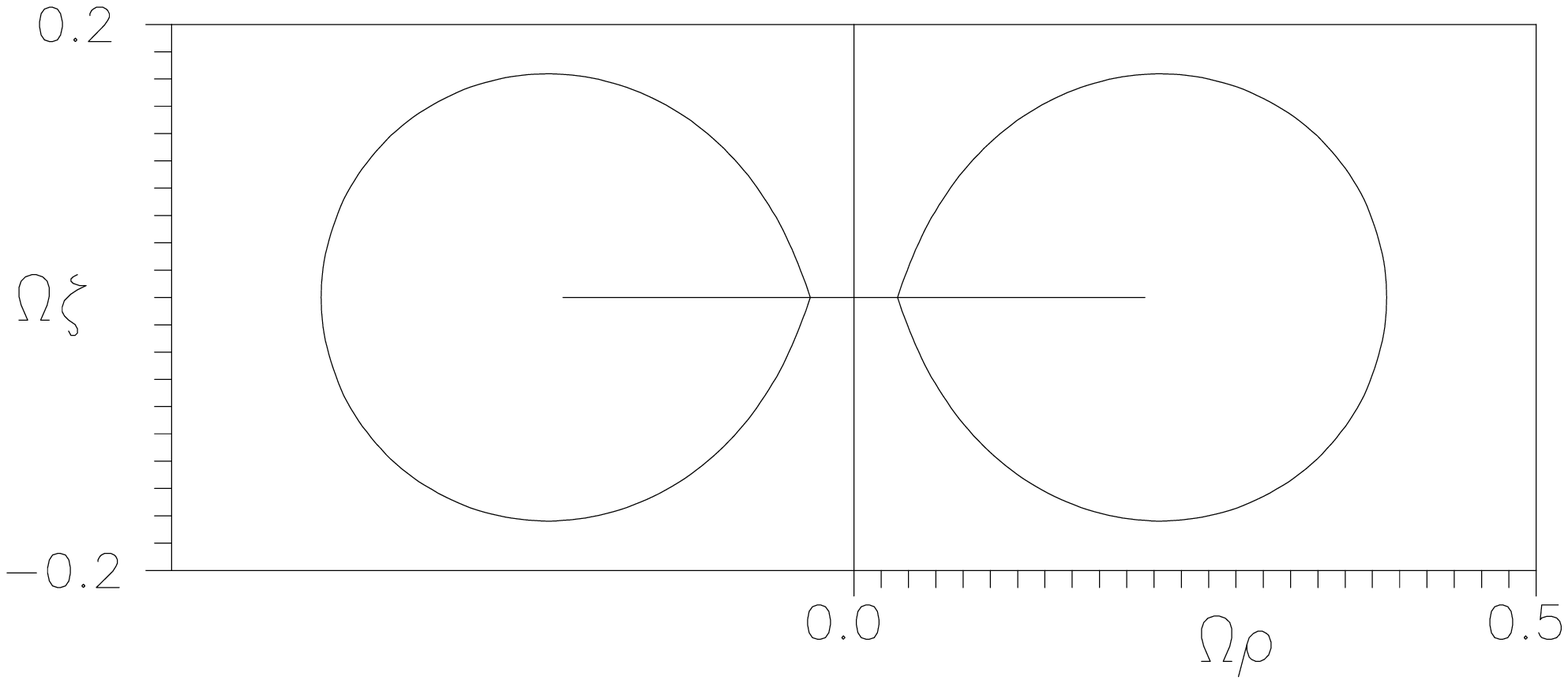}{\qquad $\mu = 3.0$}
\end{center}   
\begin{center}
\includegraphics[width=0.3\textwidth]
   {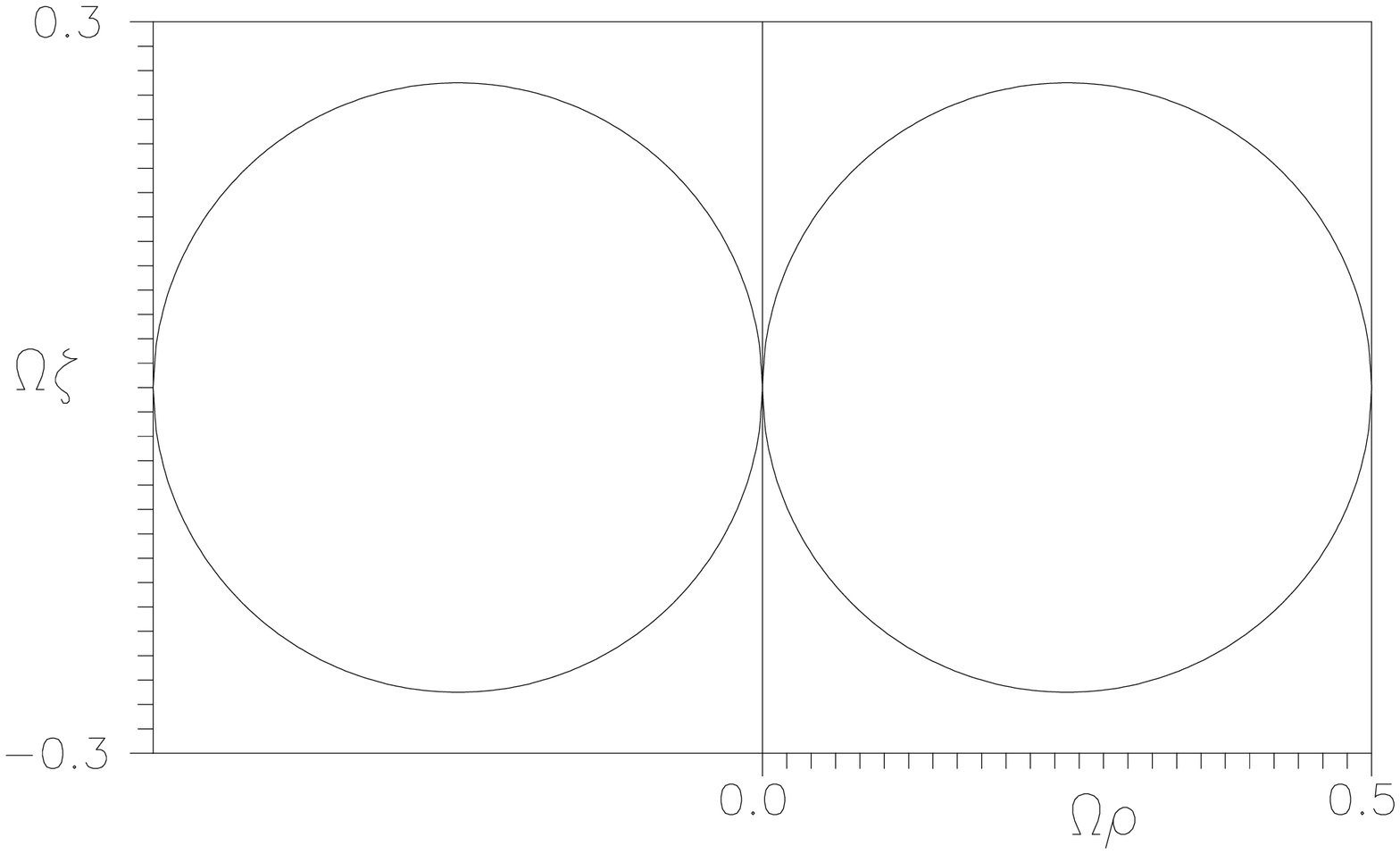}{\qquad $\mu = \mu _0$}
\end{center}

\noindent {\small{\sc Fig.~11} --- The ergoregion \cite{mk}.}

\noindent Note that the shape of the ergoregion for $\mu=3$ may be 
rediscovered in Fig.~4.

\newpage
\section{Conclusions}
In this article, we have considered the rigidly rotating disk of dust as an 
extremely flattened rotating perfect fluid body. Such controllable 
limiting procedures reducing the dimension of a body (here the thickness)
are important for the derivation of equations of motion of particles from
the dynamics of extended bodies. In our case, the motion of the 
two--dimensional mass elements is generally geodesic and independent of the
underlying perfect fluid model. In that sense, our disk of dust, like the
classical Maclaurin disk, represents a ``universal'' limit for any rigidly
rotating perfect fluid ball. There is faint hope of an explicit global
solution  for three--dimensional rotating perfect fluid sources with the 
limit (\ref{ftheta})!

Another aspect of our solution is its derivation from a boundary value
problem by means of the inverse scattering method. It may be expected that
this method will prove to be a powerful tool for the solution of other
boundary value problems for axisymmetric stationary gravitational vacuum
fields in Einstein's theory. It could also improve the insight into the
structure of the exterior fields of rotating bodies. A first step in this
direction consisted in the treatment of the reflectional symmetry of the
gravitational field with the aid of the linear problem (\ref{Lin1}),
(\ref{Lin2}) in \cite{mn1}.

Finally, as a `practical' application, the solution could be used as a 
testbed for numerical codes describing rotating star models in 
general relativity,
as, e.g., neutron stars.

There is also a formal aspect to be mentioned. Generalizing the expressions
(\ref{fint}), (\ref{jacobi}) it was possible to construct
a solution class in terms of hyperelliptic theta functions containing
an arbitrary potential function and -- depending on the genus -- an arbitrary 
number of constants \cite{mn2}. The relation of this class to the 
finite--gap class of solutions \cite{ko} requires a subtle discussion of
certain limiting procedures. We want to emphasize that it is not possible
to know {\it a priori} into which class of (known or unknown) solutions of
the Ernst equation a boundary value problem might fall. 

\subsubsection*{Acknowledgement} The authors would like to thank Olaf
Neumann, Alexander Bobenko and Olaf Richter for valuable hints concerning
the old mathematical literature on hyperelliptic functions.

\end{document}